\documentclass{elsart4}

\usepackage{amssymb}

\usepackage{graphics}
\usepackage{graphicx} 
\usepackage{amssymb}
\journal{Physica B}

\usepackage{dcolumn}
\usepackage{bm}
\usepackage{color}
\usepackage{latexsym}
\usepackage[T1]{fontenc}
\usepackage{amsmath}

\input{epsf}

\begin{document}


 \title{Vortex pinning : a probe for nanoscale disorder in iron-based superconductors
 }
 \author{C.J van der Beek$^{\mathrm a}$\corauthref{cor1},}
 \author{ S. Demirdis$^{\mathrm a}$, M. Konczykowski$^{\mathrm a}$, Y. Fasano$^{\mathrm b}$, N.R. Cejas Bolecek$^{\mathrm b}$,}
 \author{H. Pastoriza$^{\mathrm b}$, D. Colson$^{\mathrm c}$, F. Rullier-Albenque$^{\mathrm c}$}
 \address{$^{\mathrm a}$Laboratoire des Solides Irradi\'{e}s, CNRS-UMR 7642 \& CEA-DSM-IRAMIS, Ecole Polytechnique, F 91128 PALAISEAU, France } 
 \address{$^{\mathrm b}$Laboratorio de Bajas Temperaturas, Centro Atomico, Bariloche \& Instituto Balseiro, Bariloche, Argentina} 
 \address{$^{\mathrm c}$Service de Physique de l'Etat Condens\'{e}, CEA-DSM-IRAMIS, F 91198 Gif-sur-Yvette, France}
 \corauth[cor1]{Corresponding author: Laboratoire des Solides Irradi\'{e}s, Ecole Polytechnique, F 91128 Palaiseau cedex, France; Tel: +33 169 334 547; FAX:+33 16933 334 554; e-mail: kees.vanderbeek@polytechnique.edu}





\begin{abstract}
The pinning of quantized flux lines, or vortices, in the mixed state is used to quantify the effect of impurities in iron-based superconductors (IBS). Disorder at two length scales is  relevant in these materials. Strong flux pinning resulting from nm-scale heterogeneity of the superconducting properties leads to the very disordered vortex ensembles  observed in the IBS, and to the pronounced maximum in the critical current density $j_{c}$ at low magnetic fields. Disorder at the atomic scale, most likely induced by the dopant atoms, leads to ``weak collective pinning'' and a magnetic field-independent  contribution  $j_{c}^{coll}$. The latter allows one to estimate quasi-particle scattering rates.

\end{abstract}


\section{Introduction}
\label{}

Quantifying the effects of material disorder is important for the understanding of the superconducting ground state of iron-based superconductors (IBS). On the one hand, the proximity of the superconducting state to anti-ferromagnetism may lead to phase segregation that will be heavily influenced, for example, by macroscopic heterogeneity of the chemical composition. On the other hand, the superconducting ground state in these materials was proposed to have a so-called $s_{\pm}$  symmetry \cite{Mazin2008,Kuroki2008}, in which the superconducting gap not only has a different value on different Fermi surface sheets, but may also change sign from one sheet to another. Then, superconductivity is thought to be exquisitely sensitive to interband scattering \cite{Kuroki2009,Onari2009}. 

The effect of impurities is usually characterized by the scattering rate such as this can be extracted from resistivity measurements in the normal state, or from the surface resistance in the superconducting state. Another well-known but little exploited probe of microscopic disorder is the pinning of vortex lines in the superconducting mixed state. The radius of their core, of the order of the coherence length $\xi \sim 1.5$ nm, makes vortices ideal probes for impurities. The fact that the vortex density can be easily varied by orders of magnitude simply by adjusting the value of the applied magnetic field $H_{a}$ makes ``vortex matter'' sensitive to local variations of material properties on different length scales. The bulk pinning force $F_{p}$ exerted by the material disorder on the vortex ensemble inhibits the latter's motion for currents smaller than the critical current density $j_{c} = F_{p}/B$. In the following, we analyze vortex distributions and $j_{c}$-data in IBS,  show how these can be used to characterize the type of disorder, and to bracket impurity scattering rates. 

\begin{figure}[t]
\includegraphics[width=0.65\textwidth]{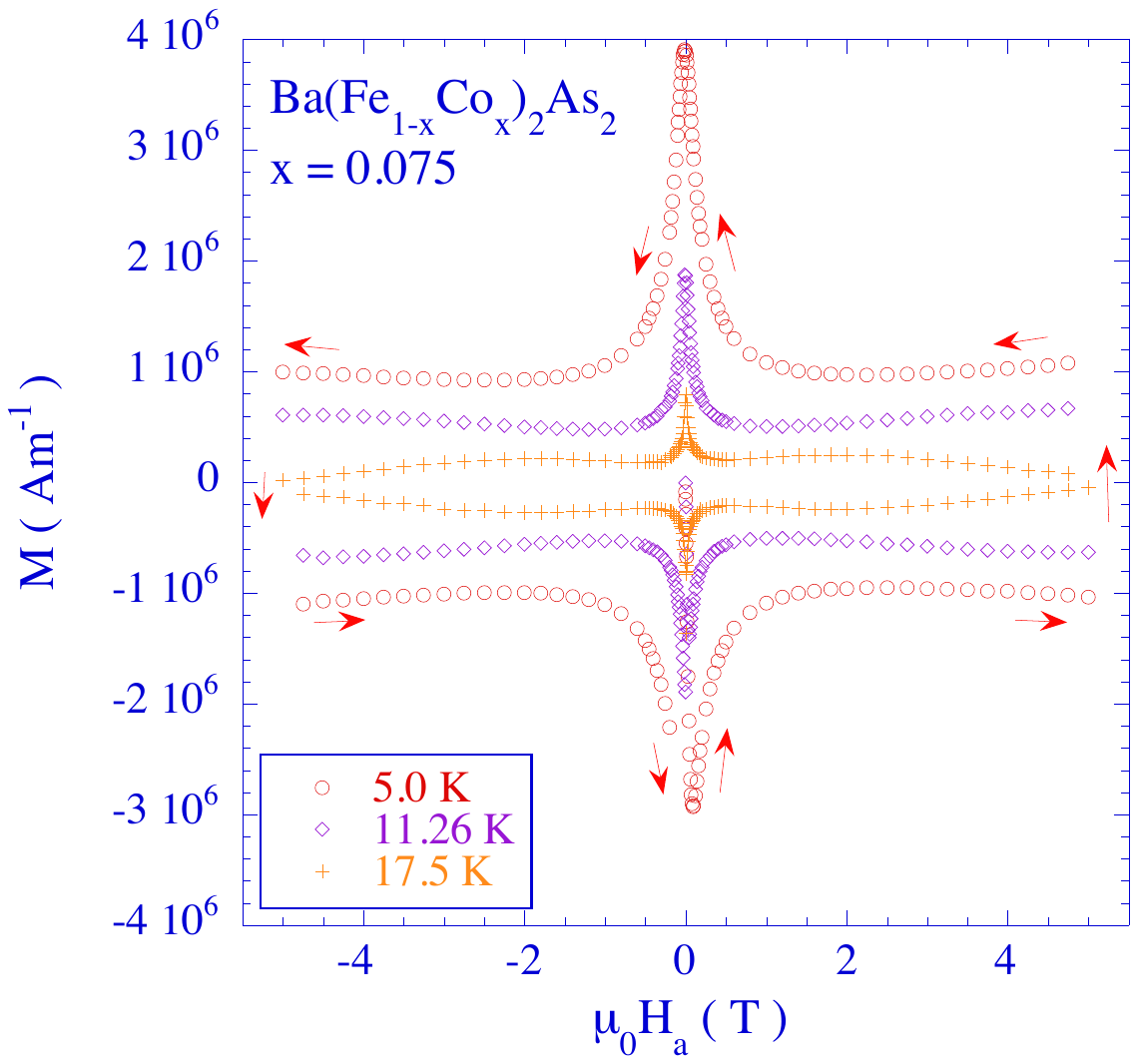}
\vspace{-6mm}
\caption{Loops of the hysteretic magnetization $M(H_{a})$ versus the magnetic field $H_{a}$, measured on a Ba(Fe$_{0.925}$Co$_{0.0.075}$)$_{2}$As$_{2}$ single crystal, at different indicated temperatures.}
\label{Loops}
\end{figure}

\section{Critical currents}

Fig.~\ref{Loops} shows hysteresis loops of the irreversible magnetization $M(H_{a})$ versus the applied magnetic field $H_{a}$, measured on a single crystal of Ba(Fe$_{0.925}$Co$_{0.075}$)$_{2}$As$_{2}$, with critical temperature $T_{c} = 23.5$ K, using a commercial Superconducting Quantum Interference Device magnetometer. As in all IBS, the hysteretic magnetic moment features a pronounced low--field maximum, superposed on a nearly field--independent contribution. A double-logarithmic plot of the critical current density as this follows from the Bean model \cite{Brandt98}, $j_{c} \sim 3 M / a$ (with $a$ the crystal width), shows that the low--field maximum amounts to a plateau $j_{c}(0)$, followed by a decrease $j_{c}(B) \sim B^{-1/2}$. Such a behavior is naturally described in terms of strong pinning by sparse, extrinsic defects of dimension greater than $\xi$, and density $n_{i} \ll \xi^{-3}$ \cite{Ovchinnikov91,vdBeek2002}. The plateau--value 
\begin{equation}
j_{c}(0)\sim (f_{p}/\sqrt{\pi}\Phi_{0}\varepsilon)(U_{p}n_{i}/\varepsilon_{0})^{1/2}
\label{eq:strong}
\end{equation}
and the decrease $j_{c}(B)\sim (f_{p}/\Phi_{0}\varepsilon)(U_{p}n_{i}/\varepsilon_{0}) (\Phi_{0}/B)^{1/2}$ can be parameterized in terms of the elementary pinning force $f_{p}$ of a single strong defect, and the ratio of the energy $U_{p}$ gained by placing a vortex on a defect to the vortex line energy $\varepsilon_{0}$ \cite{Sultan} ($\varepsilon \approx 0.4$ is the material anisotropy parameter \cite{HŠnisch2011} and $\Phi_{0} = h/2e$ is the flux quantum). Elimination of  $(U_{p}n_{i}/\varepsilon_{0})^{1/2}$ directly yields the experimental  value $f_{p} = 3 \times 10^{-13}$ N at low $T$. 
\begin{figure}[t]
\hspace{-4mm}
\includegraphics[width=0.65\textwidth]{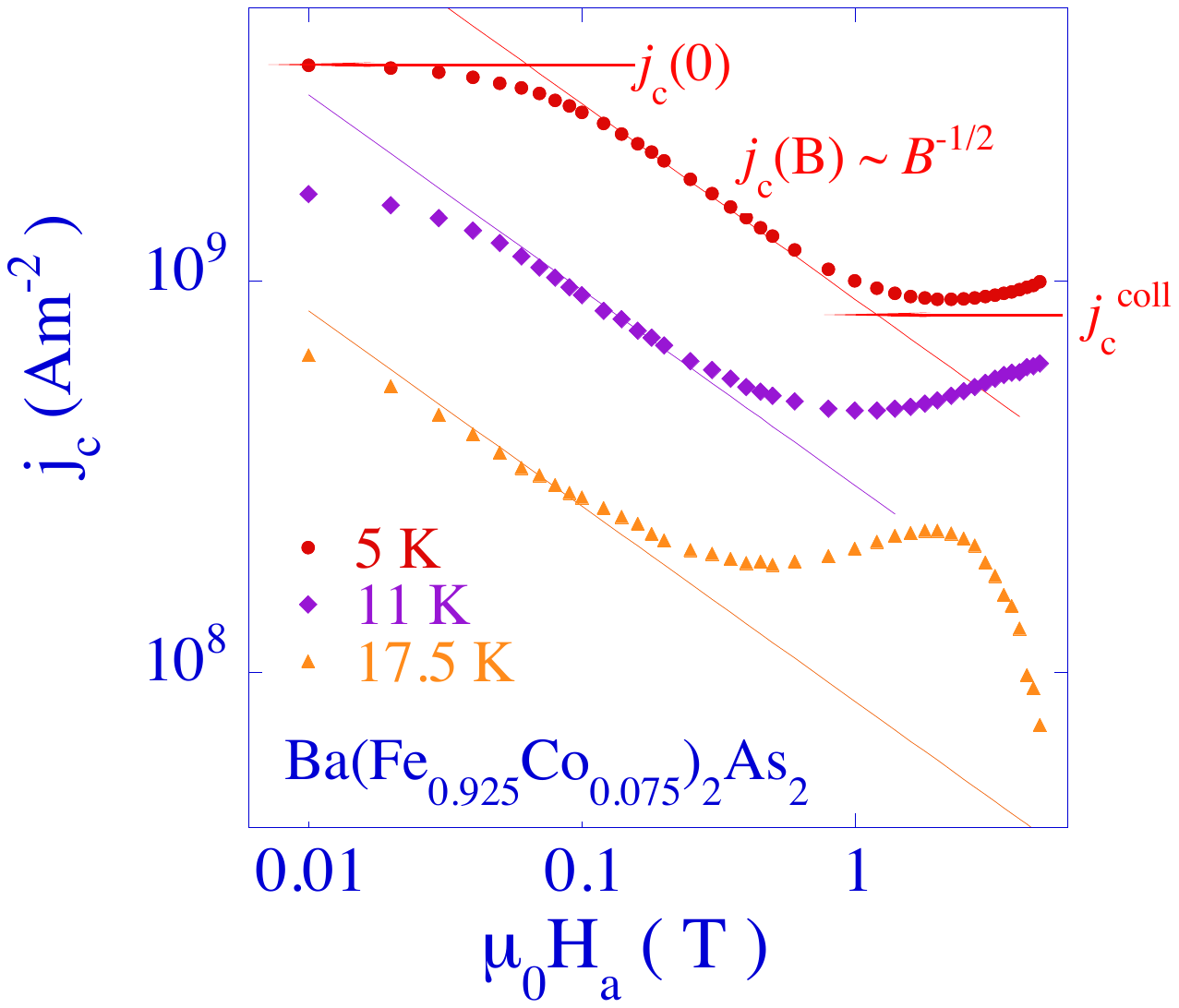}
\vspace{-6mm}
\caption{Field-dependence of the critical current density $j_{c} \sim 3 M / a$ (with $a$ the crystal width), as extracted from Fig.~\protect\ref{Loops}.}
\label{jc}
\end{figure}

\begin{figure}[b]
\hspace{-4mm}
\includegraphics[width=0.65\textwidth]{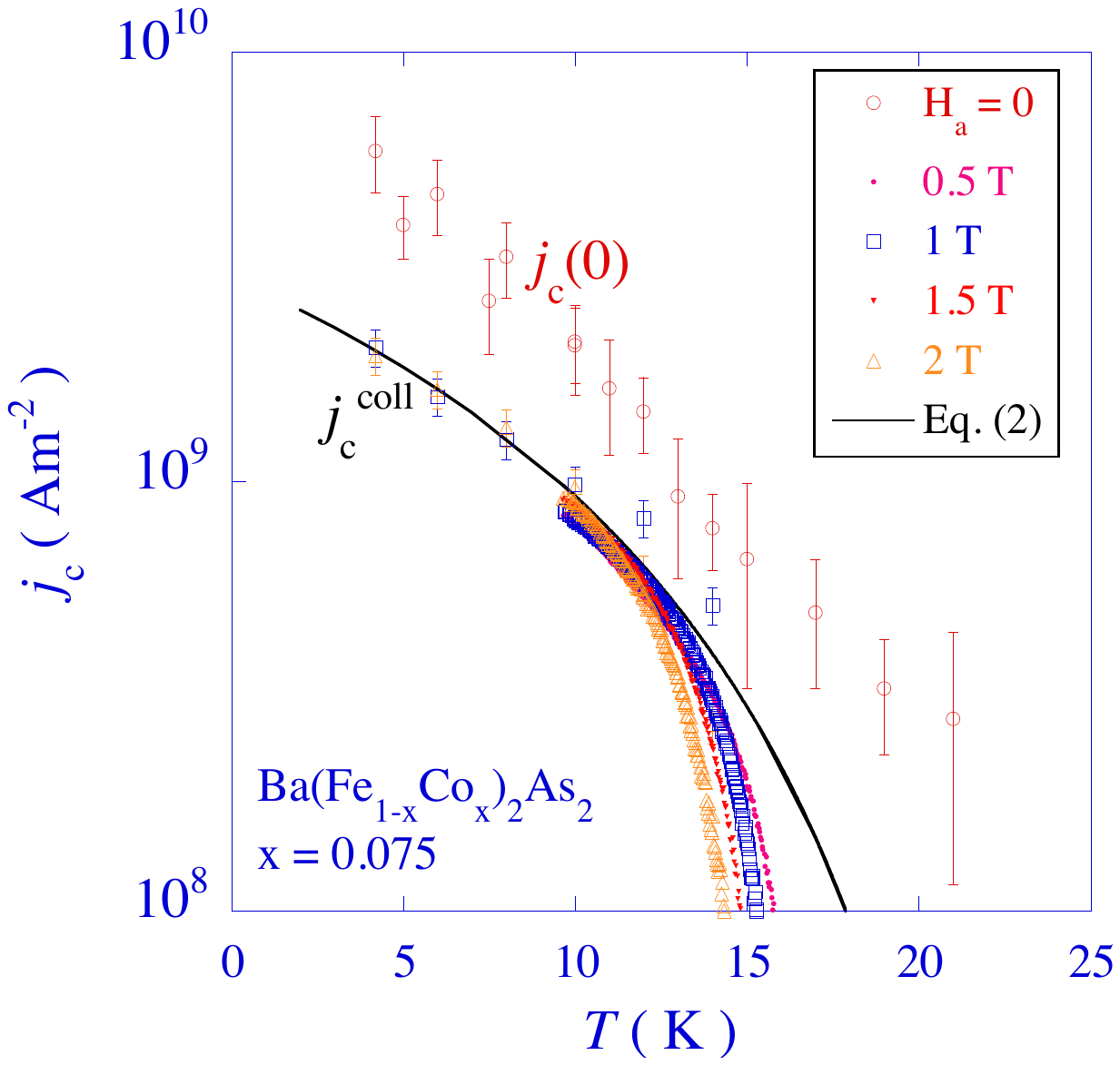}
\vspace{-6mm}
\caption{Temperature dependence of  $j_{c}$ at different applied fields. Values for $\mu_{0}H_{a} \gtrsim 1$ T are representative of $j_{c}^{coll}(T)$. The drawn line is a fit to Eq.~(\protect\ref{eq:weak}), with $\sigma_{tr} = 3$ \AA$^{2}$.}
\label{jc(T)}
\end{figure}

Further measurements were performed using the magneto-optical- \cite{Sultan,Kees1} and Hall-array techniques \cite{Kees1,LiFeAs}. The temperature dependence $j_{c}(0,T)$, and that of the field-independent contribution  $j_{c}^{coll}(T)$  observed at fields exceeding $\sim 1$ T is plotted in Fig.~\ref{jc(T)}. In the following, we first discuss the strong pinning contribution, before turning to $j_{c}^{coll}$ in Section~\ref{section:weak}.

\begin{figure}[t]
\includegraphics[width=75mm]{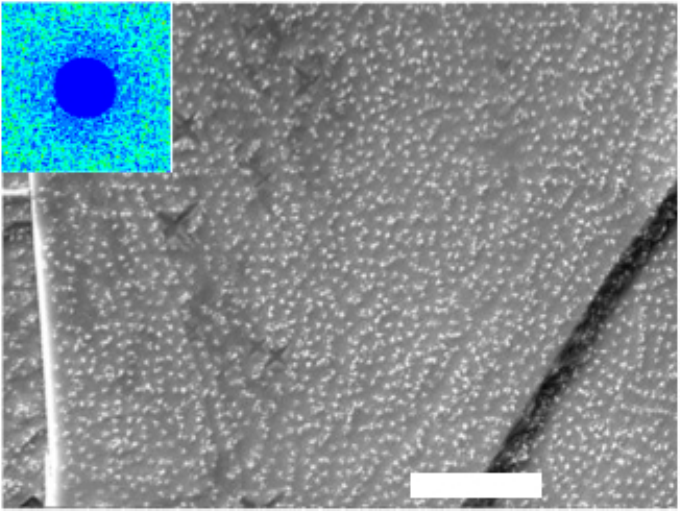}
\caption{ Bitter decoration of vortices in a Ba(Fe$_{0.9}$Co$_{0.1}$)$_{2}$As$_{2}$ single crystal, for a field $\mu_{0}H_{a} = 1$ mT applied parallel to the crystalline $c$-axis. The scale bar represents 10 $\mu$m. Inset: Fourier transform of vortex positions.}
\label{deco}
\end{figure}

\section{Heterogeneity and strong pinning}
\label{section:strong}

Fig.~\ref{deco} shows the vortex ensemble in single crystalline Ba(Fe$_{0.9}$Co$_{0.1}$)$_{2}$As$_{2}$  (with $T_{c} = 19.5$ K), at $H_{a} = 10$ G (1 mT). Vortex positions are revealed using Bitter decoration at 4.2 K, after field-cooling through the superconducting transition \cite{Sultan}. The featureless Fourier transform of the set of vortex positions indicates the absence of long--range positional- or orientational order, and the presence of large fluctuations in the nearest-neighbor distance. The very disordered vortex structure is the combined result of the narrow temperature trajectory over which it is frozen in during field-cooling and the importance of flux pinning at high $T$ \cite{Sultan}. 

\begin{figure}[b]
\vspace{-0mm}
\includegraphics[width=71mm]{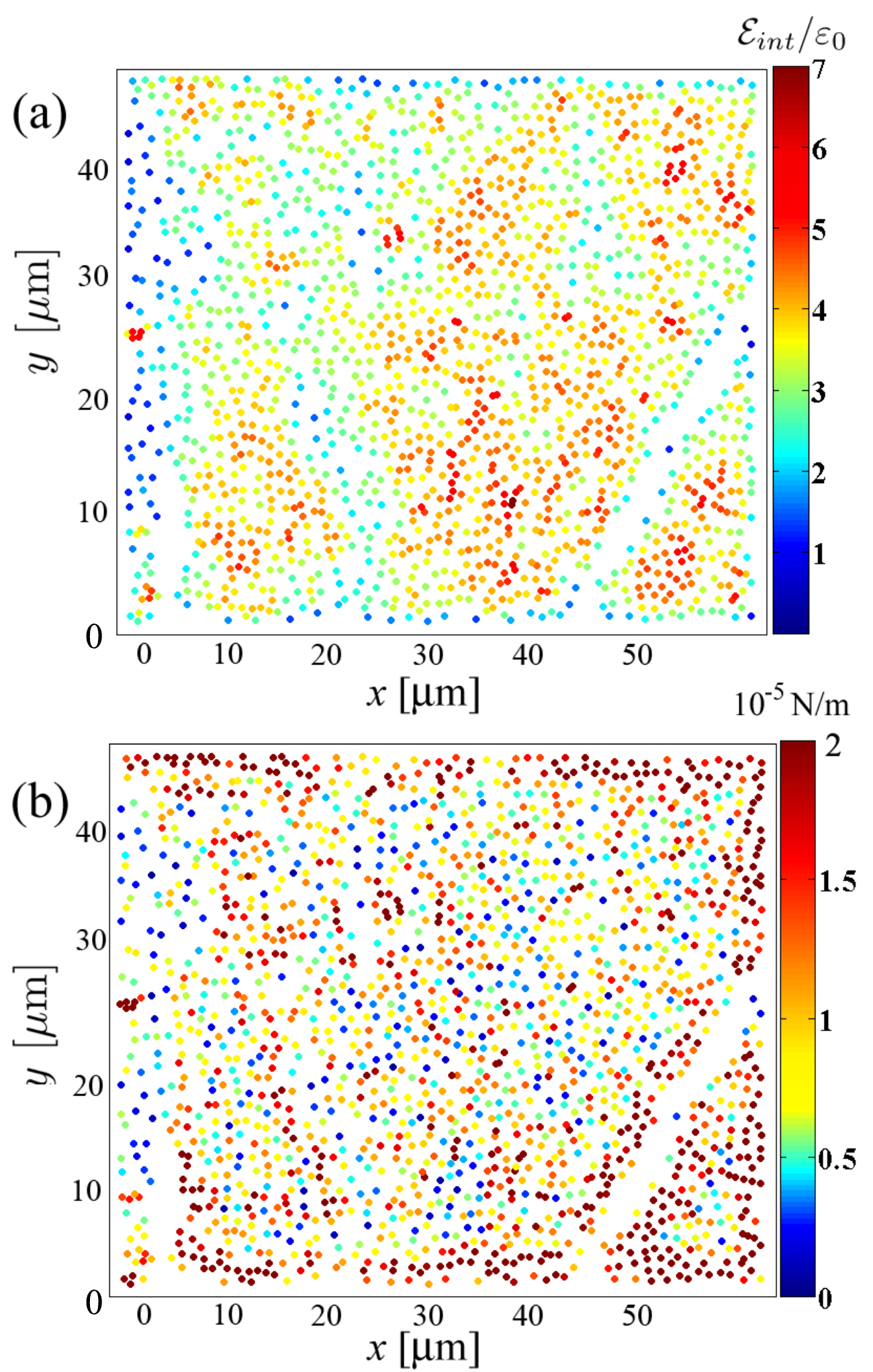}
\caption{Maps of  (a) the vortex interaction energy and  (b) the pinning force, obtained from the image in Fig.~\ref{deco}. }
\label{Maps}
\end{figure}

\begin{figure}[b]
\vspace{-0mm}
\includegraphics[width=60mm]{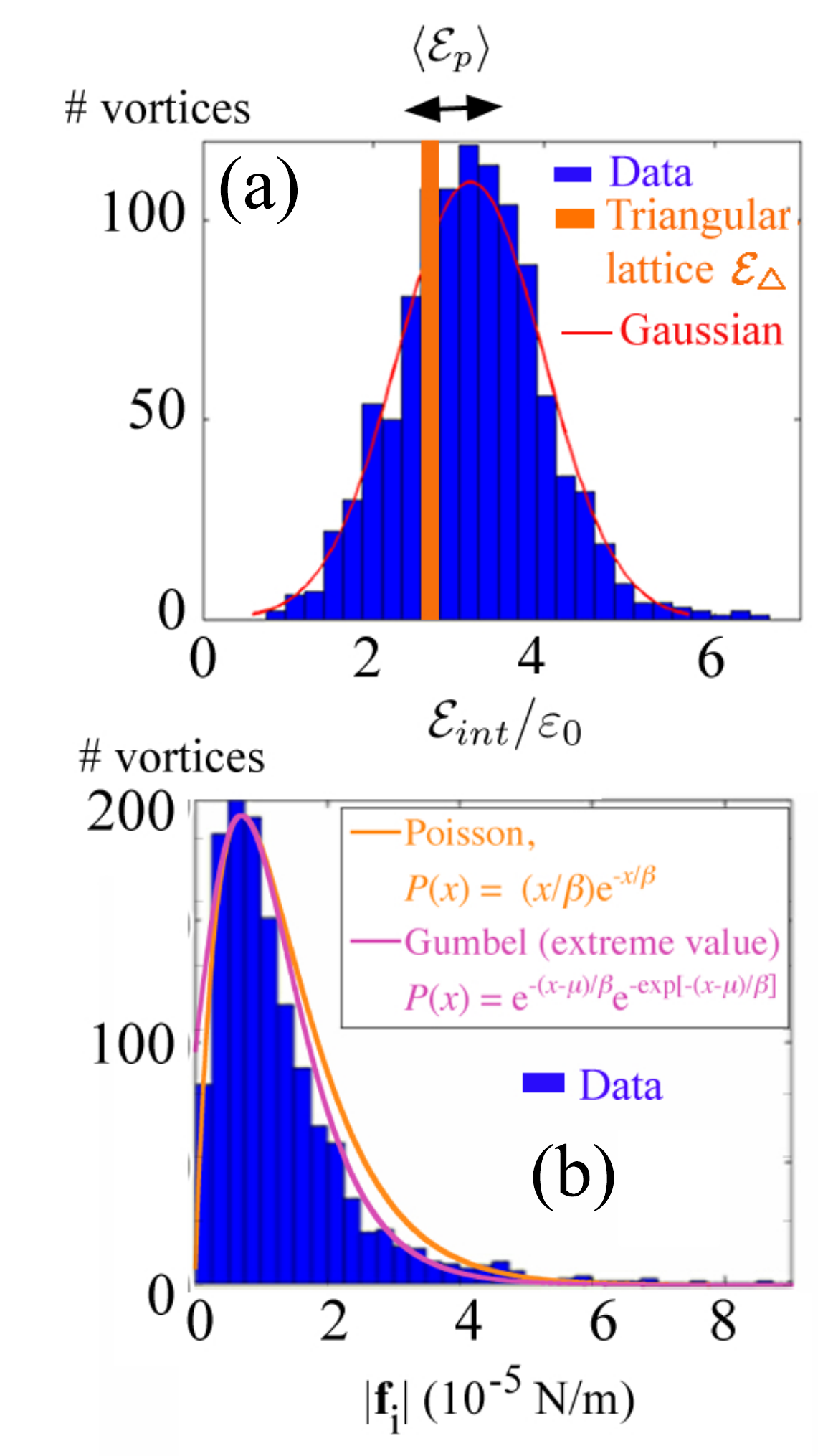}
\caption{Histograms of the vortex interaction energy (a) and of the pinning force (b) distribution of Fig.~\protect\ref{Maps}. The first is well approximated by a Gaussian, while the force distribution (b) resembles a Poisson- or extreme value distribution. The energy distribution in (a) is compared to the $\delta$-peak distribution for a (ordered) triangular vortex lattice of the same density.}
\label{Histograms}
\end{figure}

To characterize pinning, the vortex interaction energies, ${\mathcal E}_{int}^{i} =  \sum_{j} (\Phi_{0}^{2}/2\pi \mu_{0 }\lambda_{ab}^{2}) K_{0}(r_{ij}/\lambda_{ab})$, and the force to which each vortex is subjected, ${\mathbf f}_{i} =  \sum_{j} (\Phi_{0}^{2}/2\pi \mu_{0 }\lambda_{ab}^{2}) \left( {\mathbf r}_{ij} / |{\mathbf r}_{ij} | \right) K_{1}(r_{ij}/\lambda_{ab})$, were determined from the inter-vortex distance ensemble $\{r_{ij}\}$ ($\lambda_{ab}$ is the $ab$-plane penetration depth \cite{LanLuan2011}, and $K_{0}$ and $K_{1}$ are modified Bessel functions). The result is mapped out in Fig.~\ref{Maps}. The probability distribution of ${\mathcal E}_{int}^{i}$, depicted in Fig.~\ref{Histograms}a, shows that the mean interaction energy $\langle {\mathcal E}_{int}^{i} \rangle$ exceeds the value ${\mathcal E}_{\triangle}$ for vortices arranged in a triangular vortex lattice of the same (average) density. Also, the distribution is considerably broadened with respect to the regular lattice $\delta$-function. We interpret the ${\mathcal E}_{int}$--distribution as being determined by the contribution from the vortex pinning energy, with mean $\langle {\mathcal E}_{p} \rangle =  \langle {\mathcal E}_{int} \rangle - {\mathcal E}_{\triangle} \approx 0.5 \varepsilon_{0}$. This large value of  $\langle {\mathcal E}_{p} \rangle$ signifies that pinning cannot originate from well-defined defects in a homogeneous superconducting matrix, but must  be interpreted as arising from the heterogeneity of the superconducting properties of the material \cite{Sultan}.

In order to verify whether such heterogeneity can account for $j_{c}$, we turn to the distribution of $|{\mathbf f}_{i}|$. Since the vortex system is at rest, Newton's third law implies that the force map (Fig.~\ref{Maps}b) and the histogram of  $|{\mathbf f}_{i}|$ (Fig.~\ref{Histograms}b) are to be interpreted as representing the local pinning forces ${\mathbf f}_{i}$ acting on each vortex. The critical current density will be determined by the average force $\langle {\mathbf f}_{i} \rangle \sim 5 \times 10^{-6}$ Nm$^{-1}$, rather than by the most strongly pinned vortices. The effective pin density can now be estimated from the ratio of the pinning force per vortex, $\langle {\mathbf f}_{i} \rangle$, and the elementary force per pin, $f_{p} = 3\times10^{-13}$ N  determined above. This yields an average distance between effective pins of 60 nm, and a pin density $n_{i} \sim 5 \times 10^{21}$ m$^{-3} \ll \xi^{-3}$. Strong pinning in Ba(Fe$_{1-x}$Co$_{x}$)$_{2}$As$_{2}$ must therefore be due to heterogeneity on the nm scale. Substituting the obtained values in Eq.~(\ref{eq:strong}) reproduces the magnitude of $j_{c}$, for a spatial variance of the line energy $\Delta \varepsilon_{0} / \varepsilon_{0} \approx 0.05 $.


\section{Quasiparticle scattering and weak pinning} 
\label{section:weak}

The intermediate field ($\mu_{0}H_{a} \gtrsim 1$ T) plateau of constant  $j_{c} = j_{c}^{coll}$ is attributed to weak collective pinning \cite{Blatter94} by fluctuations of the dopant atom density $n_{d} \gg \xi^{-3}$ at length scales much less than the coherence length \cite{Kees}. This contribution to pinning is found in all charge-doped IBS,  as well as in Ru-doped BaFe$_{2}$As$_{2}$ and FeSe$_{1-x}$Te$_{x}$ \cite{tbp}, but not in the P-doped materials \cite{Kees}.  The magnitude and temperature dependence of $j_{c}^{coll}$ indicate \cite{Kees} that dopant atom density fluctuations are effective through the variation of the quasi-particle mean free-path they entail \cite{Blatter94,Kees,Thuneberg}. In the field-regime where vortex lines are pinned independently from their neighbors ({\em i.e.} the ``single-vortex regime'' of collective pinning \cite{Blatter94}), the resulting critical current contribution reads \cite{Kees}
\begin{equation}
j_{c}^{coll} = \frac{\Phi_{0}}{\sqrt{3}\mu_{0}\lambda_{ab}^{2}\xi}  \left[ \frac{0.01 n_{d}\sigma_{tr}^{2}}{\varepsilon \xi} \left(\frac{\xi_{0}}{\xi} \right)\right]^{2/3}
\label{eq:weak}
\end{equation}
(with $\xi_{0} = 1.35 \xi$ the $T$--independent Bardeen-Cooper-Schrieffer coherence length). Fig.~\ref{jc(T)} shows that, after insertion of a quasi-particle transport scattering cross-section $\sigma_{tr} \approx 3$ \AA$^{2}$ per Co ion, and of the $\lambda_{ab}(T)$--dependence of Ref.~\cite{LanLuan2011}, Eq.~(\ref{eq:weak}) also satisfactorily describes both the magnitude and the $T$--dependence of the intermediate--field critical current density of our Ba(Fe$_{1-x}$Co$_{x}$)$_{2}$As$_{2}$ single crystals. The extracted value of the transport cross-section can be related to the elastic quasiparticle mean-free path as $ l = (n_{d}\sigma_{tr})^{-1} \approx 26$ nm, to the scattering phase angle as $\sin \delta_{0} = (k_{F}^{2}\sigma_{tr}/2 \pi)^{1/2} \approx 0.2$, and to the quasiparticle scattering rate $\Gamma =  n_{d}[\pi N(0)]^{-1} \sin^{2} \delta_{0} \approx 3$ meV (here $N(0) \sim  mk_{F}/\pi^{2}h^{2}$ 
is the normal state Density of States and $m$ is the electronic mass) \cite{Kees}. 

\section{Summary and conclusions}
The bulk vortex pinning force $F_{p}$ and the critical current density $j_{c}$ in single crystalline Ba(Fe$_{1-x}$Co$_{x}$)$_{2}$As$_{2}$ is representative of that found in other iron-based superconductors and can be consistently described in terms of two additive contributions. At low magnetic fields, strong pinning by nm-scale variations of the vortex line energy (of the order of 5\%) is the most relevant. Such variations may arise from the inhomogeneity of the gap \cite{Massee} or from that of the superfluid density \cite{LanLuan2011} . At higher fields, the effect of quasiparticle mean-free path variations due to spatial fluctuations of the dopant atom density is dominant.

\end{document}